\documentclass{aa}
\usepackage{graphicx}
\usepackage{psfig}
\usepackage{lscape}
\usepackage{epsf} 
\begin{document}
   \title{Variability of Soft X-ray Spectral Shape: Narrow-Line Seyfert 1 Galaxies 
versus 
Broad-Line Seyfert 1 Galaxies}

   \author{Linpeng Cheng, Jianyan Wei and Yongheng Zhao
     }

   \offprints{L. P. Cheng}

   \institute{National Astronomical Observatories, Chinese Academy of Sciences, Beijing 
100012, P.R. China\\
 \email{clp@lamost.bao.ac.cn}
 }

   \date{Submitted to A\&A}

   \abstract{In order to understand how the soft X-ray spectra vary we
 present the Hardness Ratio 1 versus Count Rates (HR1-CTs) correlation 
 of 8 Narrow-line Seyfert 1 Galaxies (NLS1s) and 14 Broad-line Seyfert1 Galaxies 
(BLS1s) obtained during the ROSAT PSPC pointing observations. According to our 
criteria, six of the NLS1s show a positive HR1-CTs correlation, and seven of 
the BLS1s display an anti-correlation of HR1 versus CTs. The other 2 NLS1s 
and 7 BLS1s do not show a clear HR1-CTs correlation. From these we can see that
 the NLS1s statistically show a different spectral shape variability with flux 
change from the BLS1s: the spectra of NLS1s become harder as total flux increases while 
those of BLS1s soften. We attribute the different spectral 
variations to a strong stable 'soft excess' in NLS1s, while it is weak in BLS1s. 
For two types of objects, the power law component similarly 
becomes softer with increasing intensity. These imply that the soft excess  
originates from the Big Blue Bump and power law  emission is from Compton upscattering 
of UV or Soft X-ray photons. Our results are consistent with what is widely accepted 
that NLS1s have smaller black hole masses and higher accretion rates than BLS1s.
   
   }
   \authorrunning{L. P. Cheng, J. Y. Wei \& Y. H. Zhao } 
   \titlerunning{Spectral Shape Variability of NLS1s versus BLS1s}
   \maketitle
%

\section{Introduction}
Narrow-line Seyfert 1 galaxies (NLS1s), a peculiar group of AGNs, are characterized 
by their optical line properties: H$\beta$ FWHM is not larger than 2000 km $\rm s^{-1}$,the 
[OIII]$\lambda$5007$\rm \AA$ to H$\beta$ ratio is less than 3, and UV-optical spectrum is usually rich 
in high-ionization lines and Fe II emission multiplets (Osterbrock \& Pogge \cite{Osterbrock1985} 
). From the ROSAT All-Sky Survey it was found that about half of the AGNs 
in soft X-ray selected sample are NLS1s (Grupe \cite{Grupe1996}); in addition, Boller et al.
(\cite{Boller1996}) revealed that the soft X-ray spectra of NLS1s are systematically steeper than 
those of Broad-line Seyfert 1 galaxies (BLS1s) and that an anti-correlation exists 
between the X-ray photon index and the FWHM of the H$\beta$ line, which provides strong 
evidence for a physical link of continuum emission and the dynamics of the broad line 
region. Moreover, it was discovered that NLS1s often show stronger X-ray variance 
than BLS1s (Boller et al. \cite{Boller1996}, Leighly \cite{Leighlya}, Papadakis et al. 
\cite{Papadakis2001}). Some ROSAT (0.1-2 keV) observations revealed dramatic variability
 with giant flares, with flux 
increase by a factor of 3-5, within about a day (Boller 2000 for recent review). 
Best individual examples are IRAS 13224-3809, the flux of which varies about 2 times 
in 800 seconds (Boller et al. \cite{Boller1997}), PHL 1092 (Brandt et al. \cite{Brandt1999})
 and PKS 0558-504 (Gliozzi et al. \cite{Gliozzi2001}, Wang et al. \cite{Wang2001}). In general, 
these features are interpreted as the evidence for smaller black masses and higher accretion 
rate in NLS1s.

In the soft X-ray band below 2.0 keV, two principal components are found in the 
spectra of AGNs: a non-thermal power law emission which extends to hard X-ray band 
and a soft excess ( typically $<$ 0.5 keV ). The soft excess may be a thermal emission 
as the high energy tail of the Big Blue Bump (BBB), which is considered as the 
thermal emission from the hot accretion disk. NLS1s often contain a strong 'soft 
excess' component above extrapolated hard X-ray power law (Boller et al. \cite{Boller1996}, 
Leighly \cite{Leighlyb}). Which component dominates the variability of NLS1s? The extreme soft 
excess of NLS1s compared to BLS1s makes a first impression that their violent 
variability may be from the soft excess. If the variability is mainly from the soft 
excess, NLS1s at brighter states will show softer spectra. And if the variability 
is mainly from the power law excess, NLS1s at brighter states may show a harder spectrum,
since the soft excess dominates the soft band under 0.5 keV. Thus, the spectral 
variability of NLS1s can be used to check the origin of the variability.

 Although much attention was paid to the X-ray variability of NLS1s, most of the 
papers were on the time scales and amplitude, while seldom on the soft X-ray spectral
 shape. Spectral hardening during flux increasing was noted in three ROSAT 
observations of MARK 766 by Leighly (\cite{Leighly1996}). And Page et al. 
(\cite{Page1999}) further pointed out that the power law component of MARK 766 varied 
violently while the soft excess showed no detectable variability, and that the power 
law component became softer with increasing flux. Can the properties of MARK 766 be 
typical of NLS1s? To address the question about the variability of the soft excess 
and power law component, more NLS1s must be included. 

In this paper we report the results of the correlation between Hardness Ratio 1 and 
Count Rates (HR1-CTs) of 8 NLS1s and 14 BLS1s. BLS1s are better for analyzing the
behavior of the power law component since their soft excesses are generally much 
weaker than those of NLS1s. The 22 objects were all observed with ROSAT/PSPC in 
pointing mode. In sec. 2 we describe the observations and data reduction. The results 
are shown in Sec. 3, Sec. 4 contains our discussion; finally we take into conclusion
in sec. 5.


\section{The observations and data reduction}
The data used in this paper are from ROSAT pointed observations of individual objects 
carried out during the period from days to years, using the X-ray telescope (XRT) on
board the ROSAT observatory with PSPC on the focal plane (Tr\"{u}mper \cite{Trumper1983}). 
We selected AGNs from cross 
identification of Veron (1991)'s AGN catalogue with ROSAT pointed catalog. Using the 
ROSAT public archive of pointed observations, only sources with total X-ray photon 
counts more than 1000 are selected in order to obtain high quality X-ray spectrum. 
This yielded 214 AGNs. The data were processed for instrument corrections (such as 
the vignetting and the dead line effects) and background subsection using the EXSAS/MIDAS
 software.

The light curve for each AGN was obtained from original ROSAT observations with 
time binning of 400 seconds in three energy bands: 0.1-2.4 (total band), 0.1-0.4 
(A band), 0.5-2.0 (B band) keV. Then, we picked up 22 Seyfert 1 galaxies or quasars 
of the 214 objects by the following criteria: 1) For each source the ratio of maximal
 CTs to minimal CTs is greater than 2, which assure the range of CTs variability is 
large enough; 2) The data points is not too scarce ($>$ 5) and they distribute in one 
diagram consecutively; 3) HR1 error is small ($< 40\%$). These sources are of different
 types of AGNs, including 8 NLS1s, 14 BLS1s/QSOs.

\section{Analysis of spectral shape variability}
  \subsection{The results of HR1-CTs correlations} 
  All these X-ray count rates in 0.1-2.4 keV band were gained from original ROSAT 
observations with time binning of 400 s. In addition, 4 energy bands are shown: 
A 0.1-0.4 keV, B 0.5-2.0 keV, C 0.5-0.9 keV, D 0.9-2.0 keV. The standard hardness 
ratios, HR1 and HR2, for ROSAT-PSPC data are defined as:
   \begin{equation}
  HR1 = \frac{B-A}{B+A},  HR2 = \frac{C-D}{C+D}
   \end{equation}
"A" band is the most sensitive to the variability of the soft excess in the four 
ROSAT bands. The variability of the soft excess will show in the correlation between
 HR1 and count rates. In order to distinguish different variation trend of each object 
we fit the data through a linear formula (HR1=a+b$\times$CTs): When the slope b is a 
positive or negative value and its relative error is less than 50\%, we think it 
has a positive or negative correlation; The other instances are of random or no 
clear correlations. The results of the HR1-CTs correlation of 8 NLS1s and 14
 BLS1s/QSOs are listed in Table 1 and plotted in Fig. 1-3. The correlations are 
summarized as the following:
\begin{enumerate}
      \item For eight NLS1s in the sample, six of them, MCG-6.30.15, MARK 335, 
PG 1404+226,  MARK 766, WAS-61 and NGC 4051, show a positive HR1-CTs correlation.
 Both PG 1211+143 and TON S180 do not show any  clear variation trend.
      \item Seven of the fourteen BLS1s, NGC 7469, MARK 841, NGC 5548, NGC 3031, 
GQ COM, 3C 273.0 and IV ZW 29, demonstrate a negative HR1-CTs correlation.  The rest 
objects show random variability. \newline
\end{enumerate}
NLS1s, as a group, statistically show the trend that the spectrum becomes harder
when the count rates increase. Those BLS1s who show a clear HR1-CTs correlation 
all present an opposite trend, in other words, their soft X-ray spectra become softer with 
increasing flux.

In addition, we note that there exist some particular points in the HR1-CTs plot of 
several NLS1s such as NGC 4051 and WAS 61, which display a very high value of HR1 as
 the count rates are low, contrary to the statement above. A sudden fading of soft excess or
  a violent enhance of the power law component might interpret this hardening in the 
low state, but detailed analyses will be necessary, which is beyond the scope of this 
paper.

\begin{table*}
\caption{\bf\underline{Spectral shape variation of our selected NLS1s and BLS1s.}
          Positive: the HR1-CTS relation is positive, Negative: the HR1-CTS relation is 
negative, None: the HR1-CTS relation is random. S1n: Narrow-line Seyfert galaxy,
 Sy1.0 (1.2, 1.5, 2.0): Seyfert 1.0 (1.2, 1.5, 2.0), Q: Quasar.} 
\scriptsize
\begin{center}
\begin{tabular}{rllccccccccc} \hline
ROSAT name & Other name & RA & DEC & z & Type & HR1-CTs \\

(1RXPJ) & & (2000) & (2000)&  &   &  correlation \\ \hline

  133554$-$3417.2 & MCG-6.30.15 & 13 35 53.3 &-34 17 48 & 0.008 &  S1n & Positive\\
  000619$+$2012.4 & MARK  335   & 00 06 19.4 & 20 12 11 & 0.025 &  S1n & Positive \\
  140621$+$2223.7 & PG 1404+226 & 14 06 22.1 &  22 23 42 & 0.098 & S1n & Positive \\
  121827$+$2948.8 & MARK  766   & 12 18 26.6 &  29 48 46 & 0.012 & S1n & Positive \\
  124212$+$3317.0 & WAS 61      & 12 42 11.3 &  33 17 06 & 0.045 & S1n & Positive \\
  120310$+$4431.9 & NGC 4051    & 12 03 09.5 &  44 31 52 & 0.002 & S1n & Positive \\
  005719$-$2222.7 & TON S180    & 00 57 19.0 & -22 22 47 & 0.062 & S1n & None\\
  121417$+$1403.3 & PG 1211+143 & 12 14 17.5 &  14 03 12 & 0.085 & S1n & None\\
  230316$+$0852.2 & NGC 7469    & 23 03 15.5 &  08 52 26 & 0.017 & Sy1.5 & Negative\\
  150401$+$1026.3 & MARK  841  & 15 04 01.1 &  10 26 16 &  0.036 & Sy1.5 & Negative\\
  141759$+$2508.2 & NGC 5548    & 14 17 59.5 &  25 08 12 & 0.017 & Sy1.5 & Negative\\
  095532$+$6903.9 & NGC 3031    & 09 55 33.1 &  69 03 54 & 0.000 & Sy1.5 & Negative\\
  120441$+$2754.0 & GQ COM      & 12 04 42.0 &  27 54 11 & 0.165 & Sy1.2 & Negative\\
  122906$+$0203.2 &  3C 273.0   & 12 29 06.6 &  02 03 08 & 0.158 & Sy1.0 & Negative\\
  004215$+$4019.7 & IV ZW 29    & 00 42 16.0 &  40 19 36 & 0.102 & Sy1.0 & Negative\\
  122144$+$7518.6 & MARK  205   & 12 21 44.3 &  75 18 39 & 0.070 & Sy1.0 & None\\
  161357$+$6543.0 & MARK  876   & 16 13 57.0 &  65 43 08 & 0.129 & Sy1.0 & None\\
  132158$-$3104.2 & K08.02      & 13 21 58.1 & -31 04 10 & 0.045 & Sy1.5 & None\\
  194240$-$1019.4 & NGC 6814    & 19 42 40.5 & -10 19 24 & 0.005 & Sy1.5 & None\\
  121710$+$0711.5 & NGC 4235    & 12 17 09.8 &  07 11 29 & 0.007 & Sy1.2 & None\\
  092248$+$5120.8 & Q 0919+515  & 09 22 46.9 &  51 20 39 & 0.161 & Sy1.0 & None\\
  121920$+$0638.6 & PG 1216+069 & 12 19 20.2 &  06 38 39 & 0.334 & Q     & None\\
   \hline
\end{tabular}
\end{center}
\normalsize
\label{rafdats2.tab}
\end{table*}

\subsection{The variability: the soft excess vs. the power law component}
Which component dominates the violent variability of NLS1s? If the violent 
variability of NLS1s is mainly from the soft excess, the ROSAT A band (0.1-0.4 keV)
will be expected to change more than the B band (0.5-2.0 keV), and hence a negative 
HR1-CTs correlation is expected. The statistically positive HR1-CTs correlation of 
NLS1s shown above implies that the soft X-ray variability of these NLS1s mainly comes 
from the power law component, not from the soft excess. This is in agreement with 
the result from the detailed analysis about MARK 766 by Page et al. (\cite{Page1999}),
who found that there was no detectable change of the soft excess while the power law 
component varied. 

What is the spectral behavior of the power law component? Do NLS1s and BLS1s follow 
the same rule? The results about the BLS1s are better to answer the first question 
since the power law component generally dominated the soft X-ray emission of BLS1s. 
We have showed that the spectra of all the seven BLS1s showing clear HR1-CTs 
correlation become softer with increasing flux. For these BLS1s their power law
 components get softer when they are brighter. Our simple analysis of the HR1-CTs 
correlation of NLS1s could not reveal this question clearly and directly.  Detailed
 study about Mark 766 by Page et al. (\cite{Page1999}) revealed that the power law component 
becomes softer when it becomes brighter, while variability of the soft excess and 
neutral absorbing column is not detected. The fact that the power law component 
becomes softer as it becomes brighter also was found in the cases of NGC 4051 and
 MCG 06-30-15  (Matsuoka et al. \cite{Matsuoka1990}, Kunieda et al. \cite{Kunieda1992},
 Pounds et al. \cite{Pounds1986}, Papadakis \& Lawrence \cite{Papadakis1995}). Therefore, 
if MARK 766, NGC 4051 and MCG 06-30-15 are
 typical examples for NLS1s, the power law components of both NLS1s and BLS1s vary in 
the same way. This implies that the power law components of both NLS1s and BLS1s have
 the same physical origin. Two popular models for the power law X-ray component of radio
quiet Seyfert galaxies, both of which are based on Compton upscattering of UV or soft X-ray 
photons from an accretion disk or optically thin plasma, predict that in the variation
 of a single source the power law slope is softer as the power law flux is higher 
(Done \& Fabian \cite{Done1989}, Torricelli-Ciamponi \& Courvisier \cite{Tor1995}),
 exactly consistent with what we have found. 

As shown in MARK 766, NGC 4051 and MCG 06-30-15, it is very interesting to note that, 
although the power law component becomes softer, 
their total spectra of NLS1s get harder with increasing flux. We suggest here that 
the properties of the soft excess take charge of the spectral variability difference
 between the NLS1s and BLS1s. We take the assumption that, for both NLS1s and BLS1s, 
the power law component gets softer with increasing flux, and the soft excess remains 
almost stable during the time scale of the observations. For typical NLS1s, the A band 
is dominated by the giant thermal soft excess, and B band is dominated by the power 
law component. The increase of the power law component will change the B band relatively 
more than the A band, since the A band is dominated by the stable soft excess. Therefore,
 although the increase of the power component makes the power law component softer, it 
turns out a harder spectrum when the power law component combines with the soft excess. 
For typical BLS1s, the total spectrum is dominated by the power law component, and thus 
behaves just as the power law component. For the NLS1s and BLS1s without clear HR1-CTs 
correlation, we guess that they occupy marginal soft excess.

\section{Discussion: The X-ray properties of AGNs and the nature of the soft X-ray excess}
  \subsection{The origin of the soft excess emission}
Two kinds of models were proposed for the origin of the soft excess emission. One is 
that the soft excess is the high-energy tail of the big blue bump, which is 
considered as the thermal emission from the hot accretion disk (Ross, Fabian \& 
Mineshige \cite{Ross1992}). The other is that the soft excess is from reprocessing of the 
power law component by optically thin corona (Guilbeit \& Rees \cite{Guilbeit1988}), or by the surface
 of the accretion disc (Ross \& Fabian \cite{Ross1993}). The variability properties of the soft
 excess and the power law component are useful to check these models.

We have showed that the soft excess is more stable than the power law component. This
is consistent with the BBB origin of the soft X-ray excess. The BBB may have a longer
variability time scale than the power law X-ray emission (Done et al. \cite{Done1990}). To 
survey the reprocessing models, the corona, or the accretion disc has to be large 
enough so that there is no obvious variability of the soft excess following the 
violent variation of the power law component during the ROSAT observations. It 
implies that the accretion disc, for its small size, will not be taken as the 
reprocessing mirror. In addition, the soft excess spectrum can be well fitted by one 
or two black body spectra, and hence is consistent with the shape expected from the 
high-energy tail of a hot accretion disk (e.g. Ross, Fabian \& Mineshige \cite{Ross1992}). 
Therefore the BBB is the most promising origin of the soft excess emission.

\subsection{The X-ray properties: NLS1s versus BLS1s}
Although they were defined by Osterbrock and Pogge (\cite{Osterbrock1985}) according to their 
peculiar optical properties, NLS1s, compared to BLS1, show their most outstanding 
properties in the soft X-ray band. The first, NLS1s statistically have steeper X-ray
 spectra than BLS1s. The second, NLS1s often show shorter and stronger X-ray flux 
variability than BLS1s. The third, as found in this paper, the soft X-ray of NLS1s statistically become harder
during total flux increase, whereas BLS1s often get softer. 

The behavior of the soft excess takes charge of both the first and the third 
differences between NLS1s and BLS1s. The high $\rm L/L_{Edd}$ is popularly accepted as the 
explanation for the strong soft excess emission of NLS1s. It was first proposed by 
Pounds et al. (\cite{Pounds1995}), who noted a possible analogy of AGNs to Galactic Black Hole 
candidates whose soft X-ray spectra become steep in their high state. High accretion 
rate and thus high disk temperature cause the BBB to shift towards higher energies, 
so the high-energy tail of the BBB is apparent as the unusually strong and steep soft 
X-ray excess. Both high $\rm L/L_{Edd}$ and steep soft X-ray spectrum are also benefial to 
explain the narrow widths of the emission lines in NLS1s. If the Broad Line Region 
(BLR) scales as $\rm L^{1/2}$ and the emission clouds are gravitational virialized around the 
central massive BH, higher luminosity will keep the emission line clouds in BLR 
farther away from the center, and hence smaller velocity dispersion produces narrower 
lines. Furthermore, Wandel and Boller (\cite{Wandel1998})
suggested that AGNs with steeper X-ray spectra have stronger ionizing power, and hence 
have larger BLR and narrower emission lines than flat-spectrum AGNs with comparable 
luminosity.

The small black hole mass is considered as the driver for the violent and short-time
scale variability of NLS1s (Boller et al. \cite{Boller1996}, Laor \cite{Laor2000}). For 
the sample of Turner et al. (\cite{Turner1999}), Laor (\cite{Laor2000}) found a strong
 correlation between $\rm \sigma^2_{rms}$ and $\rm M_{BH}$, which was even stronger than
 the correlation between $\rm \sigma^2_{rms}$ and $\rm H\beta$ FWHM. The tight 
correlation between $\rm \sigma^2_{rms}$ and $\rm M_{BH}$ may be understood as the 
correlation between the variability and the size of the X-ray emitting region, which is 
supposed to scale with the mass of black hole. The results of reverberation mappings of AGNs
have made it possible to estimate the mass of black hole with the $\rm H\beta$ FWHM and
 luminosity, i.e. $\rm M_{BH} \propto FWHM^2L^\alpha$, where $\alpha$ 
has the value of 0.5 or 0.7 from different samples (Peterson \& Wandel \cite{Peterson1999},
  Wandel, Peterson \& Malkan \cite
{Wandel1999}, Kaspi et al. \cite{Kaspi2000}). 
NLS1s have the smallest black hole in the type 1 Seyferts/QSOs because of their narrowest 
$\rm H\beta$ line (see Wandel et al. \cite{Wandel1999}, and Laor \cite{Laor2000}).

\section{Conclusions}
We have presented spectral shape variability analysis of 22 Seyfert 1 galaxies by showing 
the HR1-CTs relation and found that 6 NLS1s display a positive HR1-CTs correlation 
in the sense that their spectra harden during overall flux increase, while 7 BLS1s demonstrate an 
anti-correlation, opposite to NLS1s. All the rest objects do not demonstrate any evident relation of
HR1 versus CTs. The different spectral shape variability with increasing intensity can be due to a 
strong stable soft excess below 0.5 keV in NLS1s, while it is weak in BLS1s. At the same time, both 
types of objects share a common variable power law component, which softens during its flux increase.
 These results support the 
origin of soft excess from the BBB and the power law based on Compton upscattering of UV or soft 
X-ray photons from an accretion disk or optically thin plasma. Moreover, the 
different spectral variation is consistent with what is expected if NLS1s have 
smaller Black Hole masses and higher accretion rates. Thus, the distinct spectral 
variation trends may be characteristic of BLS1 and NLS1s.    

\begin{acknowledgements}   
Acknowledgments  We thank Dr. Luo Ali for helping us with data reduction and some 
software applications. Supports under Chinese NSF (19973014), the Pandeng Project, and the 
973 Project (NKBRSF G19990754) are also gratefully acknowledged.
\end{acknowledgements}

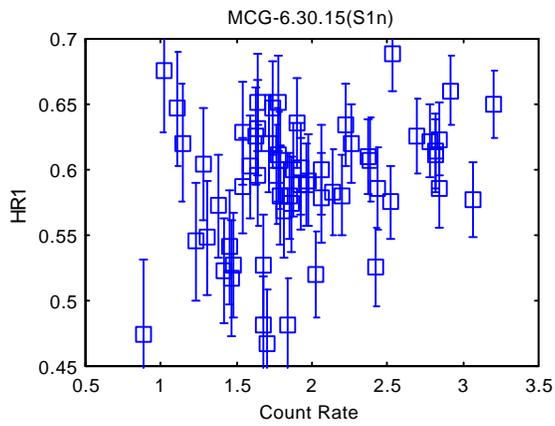
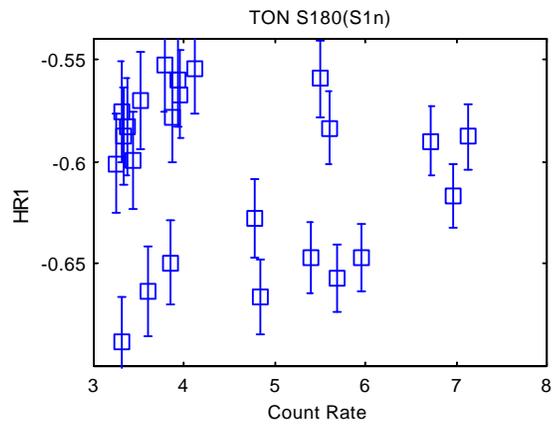
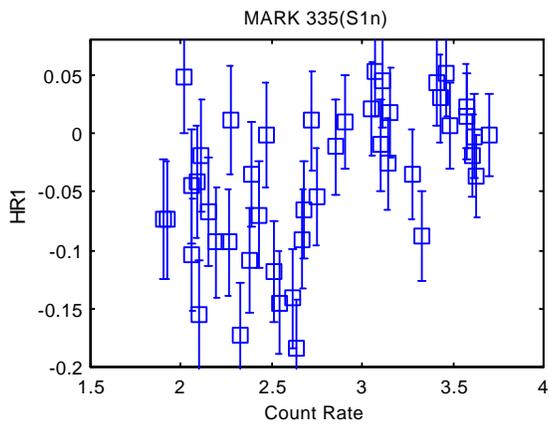
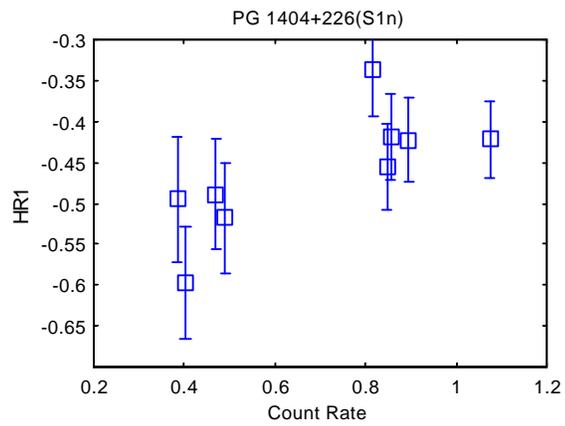
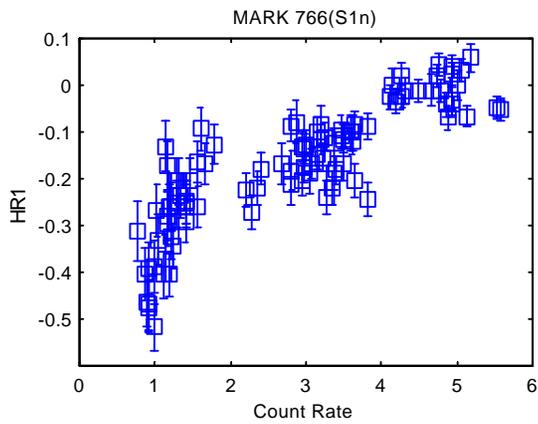
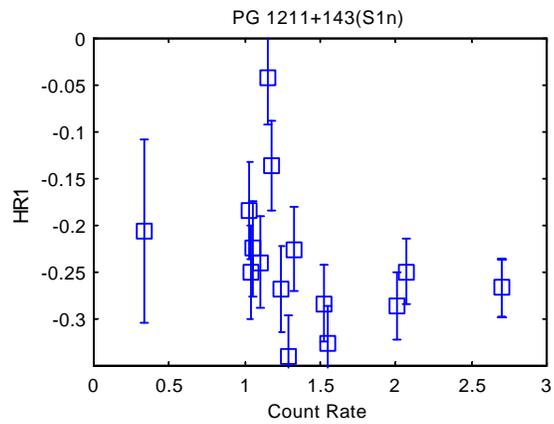
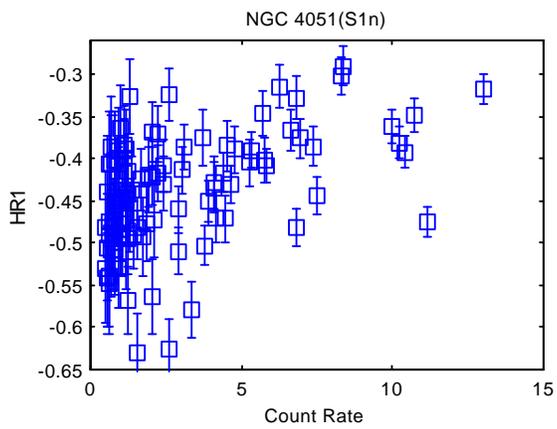
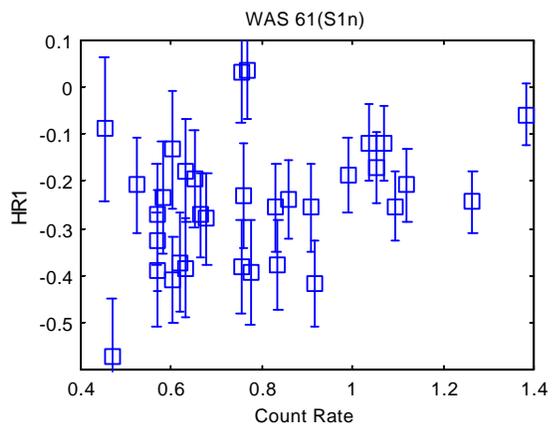
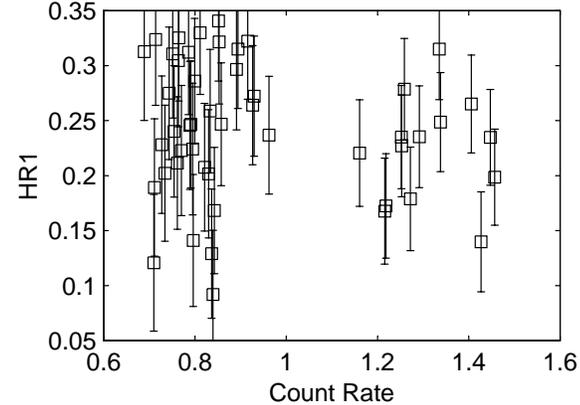
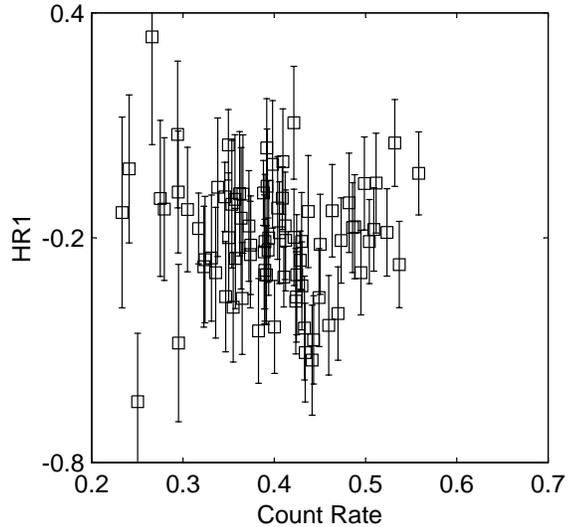
\end{document}